\documentclass[pra,twocolumn,showpacs,superscriptaddress,floatfix]{revtex4}
\usepackage{graphicx,amssymb,amsmath}

\newlength{\ldag}
\newcommand{\cra}{a^\dagger}
\newcommand{\ana}{{a^{\phantom\dagger}\hspace{-\ldag}}}
\newcommand{\crb}{b^\dagger}
\newcommand{\anb}{{b^{\phantom\dagger}\hspace{-\ldag}}}
\newcommand{\crc}{c^\dagger}
\newcommand{\anc}{{c^{\phantom\dagger}\hspace{-\ldag}}}
\newcommand{\crs}{s^\dagger}
\newcommand{\ans}{{s^{\phantom\dagger}\hspace{-\ldag}}}
\newcommand{\crt}{t^\dagger}
\newcommand{\ant}{{t^{\phantom\dagger}\hspace{-\ldag}}}

\newcommand{\tS}{\tilde{S}}

\begin{document}
\title{Concurrence in collective models}

\author{Julien Vidal}
\email{vidal@lptmc.jussieu.fr}
\affiliation{Laboratoire de Physique Th\'eorique de la Mati\`ere Condens\'ee, CNRS UMR 7600,
Universit\'e Pierre et Marie Curie, 4 Place Jussieu, 75252 Paris Cedex 05, France}
\begin{abstract}
We review the entanglement properties in collective models and their relationship with quantum phase transitions. Focusing on the concurrence, which characterizes the two-spin entanglement, we show that for first-order transitions, this quantity is singular but continuous at the transition point, contrary to the common belief. We also propose a conjecture for the concurrence of arbitrary symmetric states which connects it with a recently proposed criterion for bipartite entanglement.  
\end{abstract}


\maketitle
%
%
\section{Introduction}
\label{sec:intro}
%
%

The relationship between entanglement properties and quantum phase transitions (QPTs) has been recently the subject of many debates. An important issue is to understand how the structure of the ground state is affected by the existence of a critical point. The first answers came with the study of one-dimensional spin systems \cite{Osborne02,Osterloh02,Bose02,Syljuasen03_1,Syljuasen03_2,Latorre03,Latorre04_1,Korepin04,Keating04,Refael04,Calabrese04,Peschel04,Plenio05,Keating05,Hamma05_1,Its05,Laflorencie05,Venuti05,Qian05,Gu05,Amico06} where a strong interplay between QPTs and the variation of the entanglement with the control parameters has been in evidence. 
However, a complete classification of the possible scenarios likely to arise depending on the nature of the transitions is still lacking \cite{Wu04,Gu06}. 
In this perspective, collective models provide a good framework to investigate  these questions. Indeed, as shown in the Lipkin-Meshkov-Glick model \cite{Vidal04_1,Vidal04_2,Dusuel04_3,Dusuel05_2,Latorre05_2}, the Dicke model of superradiance \cite{Lambert04,Lambert05}, or the two-level BCS model \cite{Dusuel05_1}, the essential features met in low-dimensional systems are present in these systems which can be viewed as infinite dimensional. The main advantage of these models is that their phase diagram can be determined within a mean-field approach and that their entanglement properties, encoded in finite-size corrections, can be  computed with standard methods. 

Characterizing the entanglement of a quantum state is not an easy task and there  actually exists only a few reliable tools. 
For many-body states, one only knows how to quantify bipartite entanglement through various measures such as the block entropy (obtained after a partition of the system into two arbitrary blocks) or the concurrence \cite{Wootters98} which characterizes the entanglement between two spins after tracing out over all others.
For collective models, this latter quantity is particularly well-suited since it does not depend on the two spins selected, all spins being completely equivalent.

The goal of the present study is two fold. First, we discuss the concurrence for symmetric (permutation-invariant) states. The motivation for considering these states is that the ground states of collective models are (often) invariant under the permutation group. Simple expressions of the concurrence have already been obtained for some particular such states (parity symmetric) \cite{Wang02} but we propose here an extension of these results for a wider class. This extension leads us to conjecture a general form of the concurrence for arbitrary symmetric states that we relate to a recently proposed criterion \cite{Korbicz05_1,Korbicz06}.
Second, we analyze the behavior of the concurrence in several models in which first- and second-order transitions occur.  
The main result of this paper is that there exist situations where a discontinuity of the order parameter (associated with a first-order QPT) does not imply a jump of the concurrence at the transition point. This is a rather surprising result since the order parameter appears explicitly in the reduced density matrix from which the concurrence is extracted. 

This paper is organized as follows. In Sec. \ref{sec:conc}, we derive a simple formula for the concurrence of a class of symmetric states and propose a conjecture for arbitrary symmetric states. In Sec. \ref{sec:collec}, we introduce the collective models considered in the next sections. Section \ref{sec:bitrans} is devoted to a review of the concurrence in collective models in perpendicular magnetic field which are known to display second-order QPT. In Sec. \ref{sec:uniarbi}, we consider the most general uniaxial collective model. After discussing its phase diagram, we compute exactly the concurrence of the ground state in the thermodynamical limit as a function of the magnetic field and show that it is always continuous at the transition point. Finally, we conclude and give some perspectives in Sec. \ref{sec:conclusion}. In the Appendix, we establish the correspondence between the uniaxial model in an arbitrary field and a two-level boson model recently proposed in nuclear physics \cite{Vidal06_1} which allows a simple physical interpretation of the order parameter.

%
%
\section{A conjecture for the concurrence of symmetric states}
\label{sec:conc}
%
%
In this section, we give some simple expressions for the concurrence for symmetric states and conjecture a relationship with a recently derived bipartite entanglement criterion \cite{Korbicz05_1,Korbicz06}.

The concurrence \cite{Wootters98} is defined as follows. Let us consider a system of $N$ spins half described by a  density matrix. Let  $\rho_{i,j}$ be the reduced density matrix associated with spins $i$ and $j$ obtained by tracing out the density matrix over all spins except spins $i$ and $j$. Next, let us introduce the spin-flipped density matrix 
%
%
\begin{equation}
\tilde \rho_{i,j}=(\sigma_{y}\otimes\sigma_{y})\:\rho^{\ast}_{i,j}\: (\sigma_{y}\otimes\sigma_{y}),
\end{equation}
%
%
where $\rho^{\ast}_{i,j}$ is the complex conjugate of $\rho_{i,j}$. The
concurrence related to spins $i$ and $j$ is then defined by
%
%
\begin{equation}
C_{i,j}=\max\left\{ 0,\sqrt{\lambda_{1}}-\sqrt{\lambda_{2}}-\sqrt{\lambda_{3}}-\sqrt{\lambda_{4}}\right\},
\label{concdef}
\end{equation}
%
%
where the $\lambda_{k}$'s are the  four real positive eigenvalues  of $\rho_{i,j} \tilde{\rho}_{i,j}$, and where $\lambda_{i} \geq  \lambda_{i+1}$. If $C_{i,j}>0$, spins $i$ and $j$ are said to be entangled.

For symmetric states, it is clear that $\rho_{i,j}$ and hence $C_{i,j}$ do not depend on the specific choice of $i$ and $j$ so that we will omit these indices in the following. Now, let us try to find a simple, and physically meaningful, expression of the concurrence. As shown in Ref. \cite{Wang02}, the reduced density matrix for symmetric states in the standard basis 
$\{|\uparrow \uparrow \rangle, |\uparrow \downarrow\rangle,|\downarrow \uparrow\rangle, |\downarrow \downarrow\rangle\}$  (with  $\sigma_z |\uparrow,\downarrow \rangle= \pm |\uparrow,\downarrow \rangle$)
can generically be written as:
%
%
\begin{equation}
\rho=\left(
\begin{array}{llll}
v_{+} & x_{+}^{*} & x_{+}^{*} & u^{*} \\
x_{+} & w & y & x_{-}^{*} \\
x_{+} & y & w & x_{-}^{*} \\
u & x_{-} & x_{-} & v_{-}
\end{array}
\right)
,
\label{eq:rhored}
\end{equation}
%
%
with
%
%
\begin{eqnarray}
v_{\pm } &=&\frac{N^2-2N+4\langle S_z^2\rangle \pm 4 (N-1) \langle S_z\rangle}{4N(N-1)},  
\label{eq:aaa} \\
x_{\pm } &=&\frac{(N-1)\langle S_{+}\rangle \pm \langle
[S_{+},S_z]_{+}\rangle }{2N(N-1)},  \\
w &=&\frac{N^2-4\langle S_z^2\rangle }{4N(N-1)},  \\
y &=&\frac{ \langle S_x^2+S_y^2\rangle -N/2}{N(N-1)},  \\
u &=&\frac{\langle S_{+}^2\rangle }{N(N-1)},
\label{eq:bbb}
\end{eqnarray}
%
%
where $[A,B]_{+}=AB+BA$. Note that since for symmetric states ${\bf S}^2=\tfrac{N}{2} (\tfrac{N}{2}+1)$  one has $w=y$.

The only simple expressions of the concurrence proposed in the literature concern symmetric states that are spin-flip (or parity) symmetric, i.e., that are eigenstates of the operator $\Pi=\Pi_i \sigma_z^i$. For such states, one has $x_\pm=0$, and the concurrence is given by \cite{Wang02}
%
%
\begin{equation}
C=\Bigg\{
\begin{array}{lll}
2 \max \{ 0,|u|-y \}& {\rm if}&  2y< \sqrt{v_+ v_-}+|u| \\
2 \max \{ 0,y-\sqrt{v_+ v_-} \} & {\rm if}&   2y \geq \sqrt{v_+ v_-}+|u|
\end{array}
.
\end{equation}
%
%
An important remark is that, for real states, this result can also be written as
%
%
\begin{equation}
C= \max\{ 0,{\cal C}_y,{\cal C}_z \},
\label{eq:concparity}
\end{equation}
%
%
where,  for a given direction $n$, we have introduced the quantity 
%
%
\begin{eqnarray}
 \label{eq:defCn}
{\cal C}_n &=& {1\over 2N(N-1)} \bigg\{ N^2-4\langle S_n^2\rangle - \\
&& \sqrt{\Big[N(N-2)+4\langle S_n^2\rangle\Big]^2 - \Big[4 (N-1)\langle S_n \rangle \Big]^2}\Bigg\}. \nonumber 
\end{eqnarray}
%
%

Note that for eigenstates of $\Pi$, one has $\langle S_y \rangle=0$ so that for $n=y$, one has 
$(N-1){\cal C}_y=1-4\langle S_y^2\rangle/N$.

Before coming to our conjecture, let us also compute the concurrence for another type of states which break the parity (spin-flip) symmetry but  which have real coefficients in the standard basis  $\left\{|N/2,M\rangle \right\}$ of ${\bf S}^2$ and $S_z$.  For such states, one has 
%
%
\begin{equation}
\langle S_y\rangle=\langle[S_x,S_y]_+\rangle=\langle[S_z,S_y]_+\rangle=0,
\label{eq:Sy}
\end{equation}
%
%
so that $\rho=\rho^*$ and the characteristic polynomial associated with $\rho \tilde \rho$ factorizes:
%
%
\begin{eqnarray}
\hspace{-5mm}
P(\lambda) &=& \det (\rho \tilde \rho- \lambda), \\
                  &=& \det \big(\rho \sigma_{y}\otimes\sigma_{y}-\sqrt{\lambda} \big) \det \big(\rho \sigma_{y}\otimes\sigma_{y}+\sqrt{\lambda}\big), \\
                  &=& Q(\mu) Q(-\mu),
                  \end{eqnarray}
%
%
where we have set $\mu=\sqrt{\lambda}$. The $\lambda_i$'s of $\rho \tilde \rho$ are thus given by the squares of  the roots of $Q$.
Further, since $w=y$, one has $P(0)=0$. We thus have to analyze the roots of a third-order polynomial. 
Denoting by $\mu_i$ the three other roots of $Q$, one has
%
%
\begin{eqnarray}
{Q(\mu) \over \mu} &=& \mu^3- \mu^2 \sum_i \mu_i+ \mu \prod_{i<j} \mu_i \mu_j- \prod_i \mu_i ,\\
&=& \mu^3+2(u-y) \mu^2  + \nonumber \\
&& \mu (u^2-v_+ v_- -4u y+4x_+ x_-)   +\\
&&  2(y v_+ v_- +2 u x_+ x_- - u^2 y - v_+ x_-^2-  v_- x_+^2 ) \nonumber .
 \end{eqnarray}
%
%
The concurrence can then, under some conditions, be directly related to the $\mu_i$'s. 
Indeed, if  the $\mu_i$'s are such that $\mu_1 <0$ and $0<\mu_{2,3}<|\mu_1|$, then one has 
%
%
\begin{eqnarray}
C&=&\max\Big\{ 0,- \sum_i \mu_i \Big\} ,\\
   &=&\max\left\{ 0,2(u-y) \right\}, \\
   &=&\max\left\{ 0,{\cal C}_y \right\} \label{concCy}.
   \end{eqnarray}
%
%
The fact that the direction $y$ plays here a special role is due to the fact that the states considered have real coefficients in the basis $\left\{|N/2,M\rangle \right\}$ or, equivalently, satisfy (\ref{eq:Sy}). Finally, a close analysis of $Q$ allows one to show that the conditions $\mu_1<0$ and $0<\mu_{2,3}<|\mu_1|$ are satisfied as soon as  $u>w$. When this last condition is violated, we have no explicit form of the concurrence. 

This analysis  leads us to following conjecture: {\it the concurrence of any arbitrary symmetric state is given by
%
%
\begin{equation}
C = \max\left\{ 0,\max_n {\cal C}_n \right\},  
\label{eq:conjec1}
\end{equation}
%
%
where ${\cal C}_n$ is given by Eq. (\ref{eq:defCn})}.  We have checked numerically by choosing symmetric states with random complex coefficients in the basis that, indeed,  one always has $\max_n {\cal C}_n=\sqrt{\lambda_{1}}-\sqrt{\lambda_{2}}-\sqrt{\lambda_{3}}-\sqrt{\lambda_{4}}$. 
Although we have no proof of this conjecture, we would like to give a strong argument in its favor.  Recently, a necessary and sufficient criterion for bipartite entanglement in symmetric states has been proposed by Korbicz {\it et al.} \cite{Korbicz05_1,Korbicz06}. This criterion indicates that bipartite entanglement is present if and only if there exists a direction $n$ such that 
%
%
\begin{equation}
1-4{ \langle S_n \rangle^2 \over N^2}-4{ \langle \Delta S_n^2  \rangle \over N} >0.
\end{equation}
%
%
which is completely equivalent to  ${\cal C}_n >0$.  In this case, following Eq. (\ref{eq:conjec1}), ${\cal C}_n >0$ implies $C>0$, which is indeed a necessary and sufficient condition for bipartite entanglement. 

Now, we shall use this simple expression (\ref{eq:conjec1}) of the concurrence to characterize the entanglement of the ground state in collective models. 

%
%
 \section{Generalities about Collective models}
 \label{sec:collec}
%
%
The collective models considered here are defined as systems in which all spins mutually interact. The Hamiltonian $H$ of such systems can thus be expressed in terms of the total spin operators $S_{\alpha}=\sum_{i} \sigma_{\alpha}^{i}/2$ where the $\sigma_{\alpha}$'s are the Pauli matrices. As a direct consequence, one has 
$\left[H,{\bf S}^2 \right]=0$.
If one considers only two-spin interactions and a coupling to a magnetic field ${\bf h}$, the most general collective model Hamiltonian can be written as
%
%
\begin{equation}
  H_0=\sum_{i,j} \gamma_{i,j} S_i S_j + {\bf h.S},
    \label{eq:ham0}
\end{equation}
%
%
The interaction matrix can be diagonalized so that, after a rotation and up to some simple redefinitions, this Hamiltonian also reads
%
%
\begin{equation}
  H_1=\gamma_x S_x^2 +\gamma_y S_y^2 +\gamma_z S_z^2 + {\bf h.S},
    \label{eq:ham1}
\end{equation}
%
%
In a fixed spin sector, one further has ${\bf S^2}=S_x^2 +S_y^2 + S_z^2$, so that, finally, the most general Hamiltonian reads
%
%
\begin{equation}
  H_2=\gamma_x S_x^2 +\gamma_y S_y^2 + {\bf h.S},
    \label{eq:ham2}
\end{equation}
%
%
The most general parameter space to be investigated is thus, after normalization, four dimensional. In this work, we restrict our discussion to ferromagnetic interactions $\gamma_{x,y} <0$ for which the ground state of $H$ belongs to the symmetric space ${\bf S}^2=\tfrac{N}{2} (\tfrac{N}{2}+1)$.

For such models, a variational approach assuming a completely separable wave function is very efficient to obtain the ground state in the thermodynamical limit.  However, if one is interested in the entanglement properties, a separable state does not bring much information. As discussed in Ref. \cite{Vidal04_1}, if the concurrence of the ground state $C$ indeed vanishes in the large-$N$ limit, the rescaled concurrence $C_\mathrm{R}=(N-1) C$ has interesting properties in this limit. In the following, we will consider this latter quantity and focus on two categories of systems: the biaxial model in a transverse field and the uniaxial model in an arbitrary field. 

%
%
\section{Biaxial Model in transverse field}
\label{sec:bitrans}
%
%
Let us first consider the simplest nontrivial situation where the magnetic field is perpendicular to the interaction directions. Generically, the Hamiltonian of this system can be written as
 %
%
\begin{equation}
  H_{XY}^{\perp}=-{1 \over N} (S_x^2 + \gamma S_y^2) + h_z S_z.
    \label{eq:hamXY_perp}
\end{equation}
%
%
where $\gamma$ is the anisotropy parameter. Here, since we restrict our discussion to ferromagnetic coupling, we suppose, without loss of generality, that $0 \leq \gamma \leq 1$. We also suppose in the following that $h_z \geq 0$. As widely discussed in the literature (see, e.g., Ref. \cite{Botet83}), this system displays a second-order QPT at $h_z=1$. A good order parameter for this transition is 
%
%
\begin{equation}
m=1-4\langle S_z^2\rangle/N^2=\Bigg\{
\begin{array}{cll}
1-h_z^2 & {\rm for}&  0\leq h_z < 1, \\
0 & {\rm for}&  h_z \geq 1.
\end{array}
\label{eq:param}
\end{equation}
%
%
For a discussion of the critical exponents and the finite-size corrections associated with this transition, we refer the interested reader to Refs. \cite{Dusuel04_3,Dusuel05_2}. In the following, we simply recall the main results concerning the concurrence in these models.
%
\subsection{The anisotropic case}
%

As shown in Ref. \cite{Dusuel05_2}, the rescaled concurrence for $ 0 \leq \gamma <1$ is given by
%
%
\begin{eqnarray}
 \label{eq:concXY1}
     C_\mathrm{R}^{{(N-1)\over N} \sqrt{\gamma} \leq h_z}&=& (N-1) {\cal C}_y,\\
 \label{eq:concXY2}
   C_\mathrm{R}^{h_z\leq {(N-1)\over N} \sqrt{\gamma}}&=& (N-1) {\cal C}_z,
   \end{eqnarray}
%
%
where  ${\cal C}_n$ is defined by Eq. (\ref{eq:defCn}). These expressions (\ref{eq:concXY1}) and (\ref{eq:concXY2})  underline the role played by the special point $h_z={(N-1)\over N} \sqrt{\gamma}$ for which the two ground states can be chosen as  separable
%
%
\begin{equation}
|\psi(\theta,\phi) \rangle= \otimes^{N}_{l=1}
\Big[
\cos\left(\theta/2\right) e^{-i \phi/2}|\! \uparrow  \rangle_{l} +
\sin\left(\theta/2\right)  e^{i  \phi/2}|\! \downarrow \rangle_{l}
\Big],
\label{eq:ansatz}
\end{equation}
%
%
with $\theta= \arccos h_z$ and $\phi=0$ or $\pi$. Such a special point is often present in anisotropic models \cite{Kurmann82}. Apart from this point where $C=0$ for all $N$, the rescaled concurrence can be computed in the thermodynamical limit
%
%
\begin{eqnarray}
  C_\mathrm{R}^{1 \leq h_z}&=& 1- \sqrt{h_z-1 \over h_z- \gamma}, \label{eq:conc1} \\
  C_\mathrm{R}^{\sqrt{\gamma}\leq h_z \leq 1}&=& 1- \sqrt{1-h_z^2 \over 1- \gamma},  \label{eq:conc2} \\
  C_\mathrm{R}^{h_z\leq \sqrt{\gamma}}&=&1-\sqrt{1-\gamma \over 1-h_z^2}.  \label{eq:conc3}
\end{eqnarray}
%
%
Obviously, when $h_z$ goes to infinity, the rescaled concurrence vanishes since the ground state is, in this limit $|\psi(0,0) \rangle$ (fully polarized state in the $z$ direction). In the zero-field limit, the rescaled concurrence is nontrivial except for $\gamma=0$ where the ground states $|\psi(\pi/2,\phi=0,\pi) \rangle$ (fully polarized states in the $x$ direction) have a vanishing rescaled concurrence.

%
\subsection{The isotropic case}
%

In the isotropic case $\gamma=1$, the problem becomes simple since one further has $[H,S_z]=0$ and the eigenstates of the Hamiltonian are thus the eigenstates $\left\{|S,M\rangle \right\}$ of ${\bf S}^2$ and $S_z$. The ground state is obtained for $S=N/2$ and the following values of $M$
%
%
\begin{equation}
M_0=\Bigg\{
\begin{array}{cll}
-I(h_z N /2)& {\rm for}&  0\leq h_z < 1, \\
-N/2 & {\rm for}&  h_z \geq 1,
\end{array}
\label{eq:MGS}
\end{equation}
%
%
where $I(x)$ gives the integer part of $x$, in the following sense~:
if $x=X+\delta x$ with $X$ an integer and $\delta x\in[0,1[$,
then $I(x)=X$ for $\delta x\in[0,1/2[$ and $I(x)=X+1$ for $\delta x\in[1/2,1[$. 

As shown in Ref. \cite{Wang02}, the rescaled concurrence of the state $|N/2,M\rangle$ is:
%
%
\begin{equation}
C_\mathrm{R}= (N-1) {\cal C}_z.
 \label{eq:concXXperp}
\end{equation}
%
%
Note that these states, known as Dicke states, are parity symmetric and this expression simply comes from Eq. (\ref{eq:concparity}). In the thermodynamical limit, one thus has a discontinuity at the critical point $h_z=1$ where the rescaled concurrence jumps from 0 for $M_0=-N/2$ ($h_z>1$) to 2 for $M_0=-N/2+1$, and goes to 1 in the zero-field limit, i.e., when $M_0$ goes to 0.

These results show that the rescaled concurrence is sensitive to the existence of a QPT since it is maximum, and singular, at the critical point $h_z=1$. Further, it distinguishes between the two universality classes  $\gamma \neq 1$ and $\gamma=1$. Indeed, in the former case, $C_\mathrm{R}$ is continuous at the transition point whereas in the latter, it displays a jump despite the continuous character of the transition. To our knowledge, it is a unique example of a second-order QPT associated with a discontinuous rescaled concurrence.

%
%
\section{Uniaxial Model in arbitrary field}
\label{sec:uniarbi}
%
%
We now turn to the main contribution of this study which concerns the uniaxial model in an arbitrary field whose Hamiltonian can generically be written as
%
%
\begin{equation}
  H_X^{\perp,\parallel}=-{1 \over N} S_x^2 + h_x S_x + h_z S_z,
    \label{eq:hamX}
\end{equation}
%
%
with $h_z \geq 0$.
As discussed in the last section for $h_x=0$, this model displays a second-order transition point when varying $h_z$ with a critical point at $h_z=1$.  As we shall see, for $h_z<1$, one also faces a QPT when varying the  parallel field $h_x$. This transition is first order and occurs at $h_x=0$. Within the parametrization (\ref{eq:hamX}), it is not easy to build an order parameter characterizing this transition. However, in the Appendix, we give the correspondence between this model and a two-level boson problem recently introduced in nuclear physics \cite{Vidal06_1} which allows one to get an order parameter.
For $h_z>1$, a simple mean-field analysis similar to those presented in Ref. \cite{Dusuel05_2} also allows one to show that there is no transition when varying $h_x$. Similarly, for $h_x \neq 0$, no transition is found when varying $h_z$. 

Let us now discuss the behavior of the concurrence in the plane $(h_x,h_z)$. The ground state of $H_X^{\perp,\parallel}$ is in the maximum spin sector and has real coefficients in the basis $\left\{|N/2,M\rangle \right\}$. Further, as we shall show, one has ${\cal C}_y>0$ so that, as discussed in Sec. \ref{sec:conc}, its rescaled concurrence is given by 
%
%
\begin{equation}
C_\mathrm{R}=(N-1) {\cal C}_y.
\label{eq:concX}
\end{equation}
%
%

To compute  this quantity in the thermodynamical limit, we follow the same line as for the transverse field case \cite{Dusuel05_2}.  The first step is to introduce the Holstein-Primakoff representation of the spin operators:
%
%
       \begin{eqnarray}
         S_z&=&\cra \ana -N/2, \label{eq:HP1} \\
         S_+&=&N^{1/2}  \cra  \left(1- n_a/ N \right)^{1/2}=(S_-)^\dag, 
         \label{eq:HP2}
       \end{eqnarray}
%
%
with $S_\pm=S_x ±\pm i S_y$ and $n_a=\cra \ana$. The creation $(\cra)$ and annihilation $(\ana)$ bosonic operators satisfy the canonical commutation rules $[ \ana,\cra ]=1$. The standard way to expand these operators is to suppose 
$\langle n_a \rangle /N \ll 1$, i.e., to assume that the ground state, in the thermodynamical limit, is fully polarized in the $z$ direction. In the opposite case, one can either perform a rotation to bring the $z$-axis along the semiclassical spin direction, or shift the bosonic operators by setting $\cra=\sqrt{N} \lambda + \crb$. This latter procedure provides a macroscopic expectation value of $S_z$ which is of order $N$, and one then has $\langle n_b \rangle/N \ll 1$. We adopt this latter approach but, of course, the former is strictly equivalent.
As explained in Ref. \cite{Dusuel05_2}, we need to expand the Hamiltonian only at order $(1/N)^0$ to get the rescaled concurrence in the thermodynamical limit.
At this order, the Hamiltonian (\ref{eq:hamX}) reads
\begin{widetext}
%
%
\begin{eqnarray}
  H_X^{\perp,\parallel}&=&N {-2 \beta^2+2 h_x \beta  (1+\beta^2)
   -h_z (1-\beta^4) \over 2 (1+\beta^2)^2} + \sqrt{N} (\crb+\anb) {-2 \beta(1-\beta^2)+h_x(1-\beta^4)+2 h_z \beta(1+\beta^2) \over 2 (1+\beta^2)^{3/2}} \nonumber \\
   && -{1 \over 8 (1+\beta^2)} \Big\{ \big({\crb}^2+\anb^2 \big) \big[2-8\beta^2+
   h_x \beta (2+3\beta^2+\beta^4)\big]+2  n_b \big[2(1-6\beta^2)+h_x\beta (4 +5\beta^2+\beta^4)-4 h_z(1+\beta^2)\big] \nonumber \\
&&+   2+\beta^2\big[-4+h_x \beta(1+\beta^2)\big] \Big\}+ O(1/\sqrt{N}),
    \label{eq:hamXHP}
\end{eqnarray}
%
%
\end{widetext}
where, for convenience,  we have set $\lambda=\beta/\sqrt{1+\beta^2}$. To diagonalize $ H_X^{\perp,\parallel}$ at this order, we proceed in two steps. 
%
%
\begin{figure}[h]
  \centering
  \includegraphics[width=7cm]{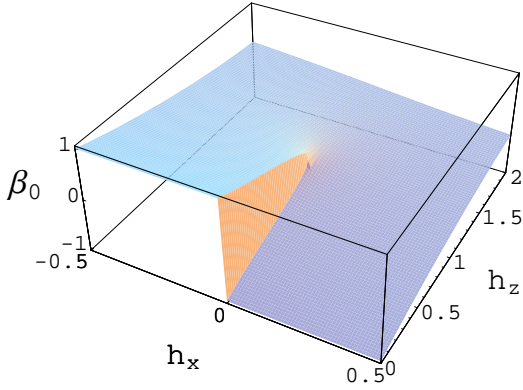}
  \caption{(Color online) $\beta_0$ as a function of $h_x$ and $h_z$. The singular behavior is clearly observed at $h_x=0$ for $h_z \leq 1$. For $h_z>1$, $\beta_0$ is a smooth function for any $h_x$.}
  \label{fig:beta0}
\end{figure}
%
%

\noindent First, we choose $\beta$ such that it minimizes the classical energy $E(\beta,h_x,h_z)$, i.e., the term proportional to $N$. This choice, $\beta_0$,  cancels the term of order $\sqrt{N}$. Then, one simply has to diagonalize a quadratic form, which is straightforward. 
A simple study of the classical energy shows that $\beta_0$ is an odd function of $h_x$ for all $h_x\neq 0$. For $h_x=0$, one has $\beta_0^2=(1-h_z)/(1+h_z)$ for $h_z \leq 1$ and $\beta_0^2=0$ otherwise. As a by-product, contrary to $\beta_0$, $\beta_0^2$ is a continuous function of the parameters $h_x$ and $h_z$. As explained in the beginning of this section, several regimes must thus be distinguished.
%
%
\begin{enumerate}

\item $h_z<1$: $\beta_0$ is discontinuous at $h_x=0$ signaling the presence of a first-order QPT on the line $(h_x=0,h_z<1)$~;

\item $h_z=1$: $\beta_0$ is continuous at $h_x=0$ but $\partial \beta_0 /  \partial h_x$ and $\partial \beta_0 /  \partial h_z$ are discontinuous functions at the point $(h_x=0,h_z=1)$.  One thus has a second-order transition point there~;

\item $h_z>1$: $\beta_0$ is a regular function  $h_x$ and $h_z$ so that no transition is expected in this region.\\

\end{enumerate}
%
%
For illustration, we plotted $\beta_0$ as a function of $h_x$ and $h_z$ in Fig. \ref{fig:beta0}. 

Once $\beta_0$ is determined we diagonalize the quadratic Hamiltonian (\ref{eq:hamXHP}) via a standard Bogoliubov transform by setting
%
\begin{equation}
      \label{eq:bogoT}
      \crc=\cosh(\Theta/2) \crb + \sinh(\Theta/2) \anb=(\anc)^\dag,
\end{equation}
%
where $\Theta$ is  such that the Hamiltonian expressed in terms of the $c$'s is diagonal. This leads us to choose
%
%
\begin{equation}
 \label{eq:Theta}
\tanh \Theta={-2+8 \beta_0^2-h_x \beta_0 (2+3\beta_0^2+\beta_0^4) \over 2(1-6\beta_0^2) 
+h_x \beta_0(4+5 \beta_0^2+\beta_0^4) -4h_z (1+\beta_0^2)}.
\end{equation}
%
%
Finally, one obtains
\begin{widetext}
%
%
\begin{eqnarray}
  H_X^{\perp,\parallel}&=&N {-2 \beta_0^2+2 h_x \beta_0  (1+\beta_0^2)
   -h_z (1-\beta_0^4) \over 2 (1+\beta_0^2)^2}  \nonumber \\
 && +  {-h_z(1+\beta_0^2)+
   \beta_0(-2\beta_0+h_x(1+\beta_0^2) \over 2(1+\beta_0^2)}+ {\Xi^{1/2}(\beta_0,h_x,h_z) \over 2}  +n_c \Xi^{1/2}(\beta_0,h_x,h_z) + 
    O(1/\sqrt{N}),
    \label{eq:hamXHP}
\end{eqnarray}
%
%
where  
%
%
\begin{equation}
\Xi(\beta_0,h_x,h_z)=\frac{
    {\left[ 2 - 12{\beta_0 }^2 - 4 h_z \left( 1 + {\beta_0 }^2 \right)  + 
        h_x \beta_0 \left( 4 + 5 {\beta_0 }^2 + {\beta_0 }^4 \right)  \right] }^2-{\left[ 2 - 8 {\beta_0 }^2 + h_x \beta_0 
          \left( 2 + 3 {\beta_0 }^2 + {\beta_0 }^4 \right)  \right] }^2}{16
    {\left( 1 + {\beta_0 }^2 \right)^2 }}.
    \end{equation}
%
%
\end{widetext}

\begin{figure}[h]
  \centering
  \includegraphics[width=7cm]{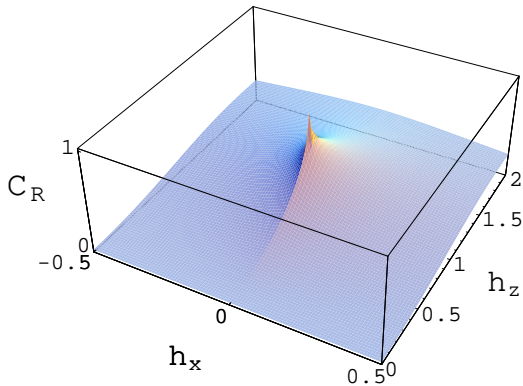}
  \caption{(Color online) Rescaled concurrence as a function of $h_x$ and $h_z$ in the thermodynamical limit. Despite the presence of a first-order transition line (for $h_x=0$ and $h_z<1$), $C_\mathrm{R}$ is a continuous function of these parameters.}
  \label{fig:conc}
\end{figure}

The ground state, at this order,  is given by the zero $c$-boson state so that we now need to write the rescaled concurrence for the ground state in terms of these bosons. Using 
Eqs. (\ref{eq:concX}), (\ref{eq:HP2}), (\ref{eq:bogoT}), and (\ref{eq:Theta}), one finds
%
%
\begin{equation}
C_\mathrm{R}=1-\frac{\sqrt{2-10 \beta_0^2 -2 h_z(1+ \beta_0^2)+h_x(3+4\beta_0^2 +\beta^4)}}{
    {1 + {\beta_0 }^2 }}.
    \label{eq:cr}
    \end{equation}
%
%
For $h_x=0$, one retrieves expressions (\ref{eq:conc1}) and (\ref{eq:conc2}) setting $\gamma=0$. The most interesting result concerns the behavior of $C_\mathrm{R}$ at the transition point $h_x=0$. Indeed, as explained above, $\beta_0^2$ is a continuous function of $h_x$ and $h_z$ so that $C_\mathrm{R}$ is also a continuous function of these parameters. In particular, when $h_z<1$,  the system undergoes a first-order QPT but the concurrence is continuous even though the order parameter has a jump.
However, it can easily be checked that for $h_x=0$ and $0<h_z \leq 1$, $\partial C_\mathrm{R} / \partial h_x$ is discontinuous but finite whereas $\partial C_\mathrm{R} / \partial h_z$ is continuous (see  Fig. \ref{fig:conc}). 
By contrast, for $h_x=0$ and $h_z=1$, $\partial C_\mathrm{R} / \partial h_x$ and  $\partial C_\mathrm{R} / \partial h_z$ are discontinuous and diverge. Finally, for $h_z>1$, $C_\mathrm{R}$ is a smooth function of $h_x$ and $h_z$ as can be seen in Fig. \ref{fig:conc}.
%
%
%
%

This example underlines the fact that the behavior of this entanglement measure does not directly inform us about the nature of the transition. It is likely that a complete classification certainly requires more information as recently discussed in Ref.  \cite{Gu06}.
This leads us to comment on the theorem proposed in Ref. \cite{Wu04} which states that a discontinuous concurrence is necessary, and sufficient, to signal a first-order QPT. Our result does not violate this theorem since one of its assumptions is not satisfied. Indeed, we face here a situation where the discontinuous elements of the reduced density matrix $\rho$ (for example $\langle S_x \rangle$) do not appear in the concurrence.

Another interesting point that has attracted much attention recently concerns the finite-size corrections to the concurrence. As already discussed in Refs. \cite{Vidal04_2,Dusuel04_3,Dusuel05_2} for $h_x=0$, the only divergence in the $1/N$ expansion of $C_\mathrm{R}$ occurs for $h_z=1$, detecting a finite-size scaling exponent for $C_\mathrm{R}$ which is 1/3. 
For $h_x=0$ and $h_z<1$, we have checked that the $1/N$ correction to  $\langle S_y^2\rangle/N$ is regular so that one expects $C_\mathrm{R}$ to have simple $1/N$ corrections in the large-$N$ limit. 
%
%
\section{Conclusion and perspectives}    
\label{sec:conclusion}
%
%
In this work, we proposed a conjecture for the concurrence of arbitrary symmetric states. This conjecture has been checked numerically and is deeply connected with a recently proposed criterion for bipartite entanglement \cite{Korbicz06}. 
Using this conjecture, we have analyzed the entanglement properties of a collective model with uniaxial interaction and arbitrary field which displays first- and second-order QPTs. For this system, we have exactly computed the rescaled concurrence $C_\mathrm{R}$ in the thermodynamical limit and in the whole parameter range. Although $C_\mathrm{R}$ is singular at the transition point, it is never discontinuous at the transition point as one could have expected.

The next step would be to consider the most general (ferromagnetic) collective system which is the biaxial model in arbitrary field given by the Hamiltonian 
%
%
\begin{equation}
  H_{XY}^{\perp,\parallel}=-{1 \over N} (S_x^2 + \gamma S_y^2) + {\bf h.S}.
    \label{eq:ham3}
\end{equation}
%
%
In this case, the parameter space to investigate is four dimensional and the eigenstates have complex coefficients in the standard basis $\left\{|N/2,M\rangle \right\}$ so that we cannot, {\it a priori}, analytically determine the direction $n$ giving the concurrence via our conjecture. 
Such a (numerical) study is beyond the scope of the present work but would certainly be of interest to understand  the competition between the interaction anisotropy and the orientation of the magnetic field. 

It would also be instructive to study the one-dimensional $XY$ model in arbitrary field which is the counterpart of 
(\ref{eq:ham3}) in finite dimensions. Even in the uniaxial (Ising) case, the behavior of the concurrence as a function of the field is an interesting issue and a comparison with the results of the present work would bring complementary  informations about the interplay between entanglement and QPT.

Finally, one may also think about using other entanglement measures such as the block entropy. For collective models in perpendicular field, it has already revealed nontrivial behavior \cite{Latorre05_2}. A complete study of the entanglement entropy for the ground state of $H_{XY}^{\perp,\parallel}$, would definitely be enriching.

\acknowledgments

I am very grateful  to S. Dusuel and J.-M. Maillard for fruitful and valuable discussions. I also thank J. I. Cirac and J. Korbicz for their comments about the conjecture proposed for the concurrence of symmetric states.

\appendix

\section{Correspondence between a two-level-boson model and the uniaxial model in arbitrary field}
\label{Casten_mapping}

In a recent paper \cite{Vidal06_1}, a simple two-level boson model has been introduced. The Hamiltonian of this model is
%
%
\begin{equation}
  H=x~  n_t-\frac{1-x}{N} ~ Q^{y} Q^{y},
  \label{eq:hamcasten}
\end{equation}
%
%
where the operators $n_t$ and $Q^{y}$ are defined as
%
%
\begin{equation}
  \label{eq:Qdef}
 n_t= \crt \ant, \hspace{0.5cm} Q^{y}=\crs \ant+\crt \ans+ y ~  \crt \ant,
\end{equation}
%
%
in terms of two species of scalar bosons $s$ and $t$, $x$ and $y$ being two independent control parameters. The total number of bosons $N=n_s + n_t$ is a conserved quantity. To map the Hamiltonian (\ref{eq:hamcasten}) onto the Hamiltonian of the uniaxial model, we use the Schwinger representation of the spin operators:
%
%
\begin{equation}
S^{+}= \crt \ans= (S^{-})^\dag, S^{z}=\frac{1}{2}(\crt \ant-\crs \ans).
\end{equation}
%
%
In terms of the spin operators, the Hamiltonian (\ref{eq:hamcasten}) then reads
%
%
\begin{equation}
  \label{eq:hamiltonian}
  H=x\Big(S_z+ \tfrac{N}{2} \Big)-\frac{1-x}{N} \Big[2 S_x +y\Big(S_z+\tfrac{N}{2} \Big) \Big]^2.
\end{equation}
%
%

Next, let us perform a rotation around the $y$ axis: 
%
%
\begin{equation}
      \label{eq:rotation}
      \left(
        \begin{array}{c}
          S_x\\ S_y\\ S_z
        \end{array}
      \right)
      =
      \left(
        \begin{array}{ccc}
          \cos\alpha & 0 & \sin\alpha \\
          0 & 1 & 0\\
          -\sin\alpha & 0 & \cos\alpha        
          \end{array}
      \right)
      \left(
        \begin{array}{c}
          \tS_x\\ \tS_y\\ \tS_z
        \end{array}
      \right),
\end{equation}
%
%
with $\tan \alpha=-y/2$. Such a rotation diagonalizes the interaction matrices so that the Hamiltonian (\ref{eq:hamcasten}) is now given by:
%
%
\begin{eqnarray}
  \label{eq:hamcastenrot}
  H&=&x \frac{N}{2}-N y^2 \frac{1-x}{4} - \tS_x^2 \: \frac{(1-x)(4+y^2)^2}{4 N} \cos^2 \alpha  + \nonumber \\
  && \tS_z \: x \cos \alpha + \tS_x \:  \frac{y(5+y^2) (x-x_c)}{2} \cos \alpha ,
\end{eqnarray}
%
where $x_c=(4+y^2)/(5+y^2)$ is known to be the transition point \cite{Vidal06_1}. The correspondence with the uniaxial model Hamiltonian (\ref{eq:hamX}) is then straighforward. Indeed,  after normalizing the Hamiltonian (\ref{eq:hamcastenrot}) to have an interaction term which is  $-\tS_x^2 /N$  the mapping reads
%
%
\begin{eqnarray}
h_x&=&\frac{y (5+y^2) (x-x_c)}{(x-1)(4+y^2)^{3/2}}, \\
h_z&=&\frac{2x}{(x-1)(4+y^2)^{3/2}}.
\end{eqnarray}
%
%
One can check that the transition point $x=x_c$ corresponds, as expected, to $h_x=0$. Let us also mention that in order to keep a ferromagnetic interaction term, one must have $x<1$ which implies some constraints on the field amplitude. 

The main advantage of the parametrization (\ref{eq:hamcasten}), in terms of the two bosons $s$ and $t$, is that the order parameter is simply $\langle n_t \rangle/N$ \cite{Vidal06_1}. \\



\end{document}